\documentstyle[aps,prl,multicol,epsf]{revtex}

\begin{document}

\draft

\title{In-plane Magnetoconductivity of Si-MOSFET's:\\
A Quantitative Comparison between Theory and Experiment.}

\author{S. A. Vitkalov$^{(a)}$, K. James$^{(a)}$,
B. N. Narozhny$^{(b)}$, M. P. Sarachik$^{(a)}$, and
T. M. Klapwijk$^{(c)}$}

\address{
$^{(a)}$Physics Department, City College of the City University of
New York, New York, New York, 10031 \\
$^{(b)}$Department of Physics, Brookhaven
National Laboratory, Upton, NY 11973-5000\\
$^{(c)}$Department of Applied Physics, Delft University of
Technology, 2628 CJ Delft, The Netherlands}

\date{\today}

\maketitle

\begin{abstract}
For densities above $n=1.6 \times 10^{11}$ cm$^{-2}$ in the
strongly interacting system of electrons in two-dimensional silicon
inversion layers, excellent agreement between experiment and the
theory of Zala, Narozhny and Aleiner is obtained for the response of
the conductivity to a magnetic field applied parallel to the plane of
the electrons.  However, the Fermi liquid parameter $F_0^\sigma(n)$
and the valley splitting $\Delta_V(n)$ obtained from fits to the
magnetoconductivity, although providing qualitatively correct behavior
(including sign), do not yield quantitative agreement with the
temperature dependence of the conductivity in zero magnetic field.
Our results suggest the existence of additional scattering processes
not included in the theory in its present form.
\end{abstract}

\pacs{ PACS No:  }

\begin{multicols}{2}

Two-dimensional systems of electrons \cite{krav,papadakis,hanein2} and
holes \cite{coleridge,hanein,simmons,savch,coler} have been the focus
of a great deal of attention during the last few years
\cite{exr}. Their resistance exhibits metallic temperature dependence
above a critical electron density $n_c$, and a strong dependence on
in-plane magnetic field \cite{ferro}. Recent theory of Zala, Narozhny
and Aleiner (ZNA) \cite{int,mpf} attempts to account for the behavior
of the conductivity of these strongly interacting electron systems
using a Fermi liquid approach. Within this theory, the temperature and
field dependence of the correction to the conductivity due to
electron-electron interaction arises from the suppression of quantum
interference between different electron trajectories (i.e. coherent
scattering off Friedel oscillations). In this paper we present a
detailed quantitative comparison of the ZNA theory \cite{int,mpf} with
the in-plane magnetoconductivity of the 2D electron system in
Si-MOSFET's measured at different temperatures and different electrons
densities. Excellent agreement between theory and experiment is
obtained for the magnetic field dependence, allowing the determination
of the Fermi liquid constant $F_0^\sigma$ and valley splitting energy
$\Delta_V$. However, the temperature dependence of the conductivity at
zero magnetic field is considerably stronger than predicted
theoretically (for the same values of $F_0^\sigma$ and
$\Delta_V$). Our results suggest the presence of additional scattering
processes not included in the theory \cite{int,mpf}.

We begin with a very brief review of the theory. The conductivity of a
disordered electron system (characterized by the large dimensionless
conductance $g=2\pi\hbar/e^2R_\Box\gg 1$; $R_\Box$ is the sheet
resistance)  can be expressed as follows\cite{aar}:

\begin{equation}
\sigma = \sigma_D + \delta\sigma_{WL} + \delta\sigma_{ee}
+ {\cal O}\left(\frac{1}{g}\right)
\label{sigma}.
\end{equation}

\noindent
Here the Drude conductivity is $\sigma_D = n e^2 \tau / m^*$,
$n$ is the electron density, $\tau$ is the elastic mean free time
and $m^*$ is the
effective mass. The weak localization correction \cite{aar}

\begin{equation}
\delta\sigma_{WL} = - \frac{e^2}{2\pi^2\hbar} \ln
\frac{\tau_\varphi}{\tau}
\label{wl}
\end{equation}

\noindent
depends on external parameters through the dephasing time
$\tau_\varphi$ (see Refs.~\onlinecite{aar,aag,nza} for
details). In zero magnetic field the resulting temperature dependence
is logarithmic, unlike the correction due to electron-electron
interaction $\delta\sigma_{ee}$, which is linear in temperature in the
intermediate temperature regime (see Ref.~\onlinecite{int} and below).
Based on Eq.~(\ref{wl}) and the available experimental data
\cite{Brunth}, one can estimate the value of the weak
localization correction (\ref{wl}).  For electron density $n=2.47\times
10^{11}$ cm$^{-2}$ used in this work (see below), Eq.~(\ref{wl}) yields
$\delta\sigma_{WL}= -0.42 \times e^2/h$ at $T=0.25$ K and
$\delta\sigma_{WL}= -0.22 \times e^2/h$ at $T=1$ K. These values are much
smaller than typical variations of the conductivity with in-plane
magnetic field $H$ (see Fig.~\ref{fig1}). We therefore neglect the weak
localization correction in the present paper.

The correction term due to interactions, $\delta\sigma_{ee}$
\cite{aar}, was recently calculated by ZNA
\cite{int} for all temperatures smaller than the Fermi energy.
Physically, this correction  arises due to coherent scattering off
Friedel oscillations. For $H=0$ the result of
Ref.~\onlinecite{int} consists of the charge (singlet) channel
contribution

\begin{mathletters}
\label{all}
\begin{equation}
\delta\sigma_C =
\frac{e^2}{\pi\hbar} \frac{T\tau}{\hbar}
\left[ 1 -\frac{3}{8}f(T\tau)\right]
-\frac{e^2}{2\pi^2\hbar}\ln\frac{E_F}{T},
\label{singlet}
\end{equation}

\noindent
and the triplet channel contribution (i.e. the contribution of
spin-exchange processes)

\begin{eqnarray}
\delta\sigma_T = &&
\frac{F_0^\sigma}{(1+F_0^\sigma)}
\frac{e^2}{\pi\hbar}\frac{T\tau}{\hbar}
\left[ 1 -\frac{3}{8}t(T\tau; F_0^\sigma)\right]
\nonumber\\
&&
\nonumber\\
&&
-\left(1-\frac{1}{F_0^\sigma}
\ln(1+F_0^\sigma)\right)
\frac{e^2}{2\pi^2\hbar}\ln\frac{E_F}{T}.
\label{tc}
\end{eqnarray}

\noindent
Note that the latter must be multiplied by the number of channels ($3$
in the absence of valley degeneracy and the Zeeman splitting
$E_z=\mu g_0 H$; the Lande $g$-factor $g_0=2$);

\[
\delta\sigma_{ee} = \delta\sigma_C + 3 \delta\sigma_T.
\]

\noindent
Magnetic field freezes those channels that correspond to non-zero
total spin components, giving rise to a dependence of the conductivity
on magnetic field \cite{mpf}:

\begin{eqnarray}
\label{mr}
&&\delta\sigma(E_z) = \delta\sigma_{ee}(E_z) - \delta\sigma_{ee}(0) =
\frac{e^2}{\pi\hbar}
\left[
\frac{2F_0^\sigma}{(1+F_0^\sigma)}\frac{T\tau}{\hbar}
\right.
\\
&&
\nonumber\\
&&
\times K_{b}\left(\frac{E_z}{2T}, F_0^\sigma\right)
+ \left. K_{d}\left(\frac{E_z}{2\pi T}, F_0^\sigma\right)
+ m(E_z\tau, T\tau; F_0^\sigma) \right].
\nonumber
\end{eqnarray}

\noindent
Here the ballistic (i.e. dominant for $T\tau\gtrsim 1$) contribution
to Eq.~(\ref{mr}) is

\begin{eqnarray}
K_{b}(x, F_0^\sigma) =  x\coth x -1
+ K_{2}(x, F_0^\sigma),
\label{b1}
\end{eqnarray}

\end{mathletters}
\noindent
In the above expressions the dimensionless functions $f(x)$, $t(x,
F_0^\sigma)$, and $m(y, x; F_0^\sigma)$ describe the crossover between
the diffusive ($T\tau\ll 1$) and ballistic regimes. These functions
decay smoothly from unity to zero and change the value of
$\delta\sigma_{ee}$ by only few per cent outside the crossover region.
The details regarding these functions, as well as explicit expressions for
$K_{2}(x, F_0^\sigma)$ and the diffusive part of the
magnetoconductivity $K_{d}(E_z / 2\pi T, F_0^\sigma)$, can be found in
Refs.~\cite{int,mpf}.

The purpose of this paper is to present a detailed quantitative
comparison of the ZNA theory \cite{int,mpf} with the results of
transport measurements in Si-MOSFET's. Note that this theory (as
described in Refs.~\onlinecite{int,mpf} and above) considers an
idealized 2DEG without taking into account material-dependent details.
For 2D electron systems in Si-MOSFET's an important additional feature
is the valley degeneracy \cite{afs}. If we neglect inter-valley
scattering (in other words, assume that the valley index is a good
quantum number) and assume that the two valleys are degenerate, then
we observe that the valley index can be treated as an pseudo-spin. In
this case electrons have not one but two ``spin'' indices, and the
total number of scattering channels is thus $16$ (as opposed to $4$ in
the usual case considered in Ref.~\onlinecite{mpf}). In reality, the
valleys are split \cite{afs} with an energy splitting $\Delta_V$. This
splitting is similar to the effect of the Zeeman field, since it only
shifts energies of the electronic states without affecting the
corresponding wave functions.

Taking into account the valley degeneracy, we can write the
interaction correction to the conductivity in the following form:

\begin{eqnarray}
\delta\sigma_{ee} && = \delta\sigma_C + 15 \delta\sigma_T
+2 \delta\sigma(E_z)
\label{ee}
\\
&&
\nonumber\\
&&
+ 2 \delta\sigma(\Delta_V)
+ \delta\sigma(E_z + \Delta_V) + \delta\sigma(E_z - \Delta_V),
\nonumber
\end{eqnarray}

\noindent
where all the terms in the second line are defined by Eq.~(\ref{mr}).
Equation (\ref{ee}) is the main theoretical result which we intend to
test against the data. Our strategy is the following. The theoretical
expressions Eqs.~(\ref{all}), (\ref{ee}) contain three
phenomenological parameters: the Fermi liquid constant $F_0^\sigma$,
the elastic mean-free time $\tau$, and the valley splitting
$\Delta_V$.  For purposes of comparison with experiment, the
parameters $F_0^\sigma$ and $\Delta_v$ are free fitting parameters.
Both will be determined below from the magnetic field dependence of
the conductivity.  On the other hand, one has much less freedom in the
determination of the parameter $\tau$.  Since the theory calculates
temperature corrections to the impurity scattering time, the value of
$\tau$ in all theoretical expressions is the scattering time that the
system would have at $T=0$ in the absence of quantum corrections.
Therefore $\tau$ must be determined from the temperature dependence of
the conductivity: one has to extrapolate the linear temperature
dependence from intermediate temperatures to $T=0$ and use the
$y$-intercept as the zero temperature value of the Drude conductivity
(thus neglecting the logarithmic corrections; therefore this value
does not coincide with the actual measured residual conductivity of
the system).

The theory described above is valid \cite{int} when the dimensionless
conductance of the system is large $g\gg 1$ and the system is not too
close to the Stoner instability $T < (1+F_0^\sigma)^2E_F$.  Also, the
magnetic field dependence Eq.~(\ref{ee}) is valid for relatively weak
fields, well below the full polarization of the electron system. The
comparison between experiment and theory presented in this paper is
restricted to a range of densities and temperatures well within these
limits (see below and Figs.~\ref{fig1},~\ref{fig2}).

The remainder of the paper is devoted to a direct comparison of the
data with Eq.~(\ref{ee}).  Measurements were taken on two Si-MOSFET's
with mobility $\mu \approx 25,000\;$V/(cm$^2$s) at $0.3$ K. Data were
obtained using standard four-terminal AC techniques on samples with
split-gate geometry to $12$ T at City College and in fields up to $20$
T at the National High Magnetic Field Laboratory \cite{vitkalov}.  The
dimensions of the measured portion of the 2DEG are $120\times 50$
$\mu$m$^2$. Measurements were taken at temperatures down to $0.25$ K
in the linear regime using small currents of about $1-2$ nA to prevent
overheating the electrons.

We first consider a comparison of the magnetoconductivity data with
the ZNA theory. In Fig.~\ref{fig1} we show the longitudinal
conductivity $\sigma_{xx}$ as a function of magnetic field $H_{||}$
applied parallel to the plane of the Si-MOSFET for electron density
$n=2.47\times10^{11}$ cm$^{-2}$ at different temperatures. In
agreement with earlier results \cite{ear}, there is a substantial
decrease of the conductivity with increasing magnetic field.  The grey
lines are obtained by direct evaluation of Eq.~(\ref{ee}) using
$F_0^\sigma$ and $\Delta_V$ as fitting parameters; the value of
$\Delta_V$ was constrained to be approximately independent of
temperature, as expected within the theory.  The theoretical results are
shifted vertically to match the experimental values of the
conductivities.  The resulting values of $F_0^\sigma$ and
$\Delta_V$ are shown in the inset to Fig.~\ref{fig1}\cite{accuracy}.  The
determination of the value of $\tau$ used in this calculation is more
subtle.  As pointed out earlier, the Drude conductivity was determined
from the temperature dependence shown in Fig.~\ref{fig3} by extrapolating
the linear part of the curves to zero temperature. In order to extract the
value of $\tau$ one needs to know the effective mass of the
electrons. One can measure the product $m^*g^*$ \cite{gm}
experimentally (by analyzing Shubnikov-de Haas data, for
instance). The renormalized value of the Lande $g$-factor is related
to the same \cite{note} Fermi liquid constant $g^*=g_0/(1+F_0^\sigma)$. Using
these relations self-consistently, we obtain the value of $\tau$ for
each electron density (with $F_0^\sigma$ taken at $T=0.25$ K). Once
determined, $\tau$ was assumed to be independent of temperature. For
the density $n=2.47 \times 10^{11}$ we find $\tau = 5.28$ ps.

{
\narrowtext
\begin{figure}[ht]
\epsfxsize=6.7 cm
\centerline{\epsfbox{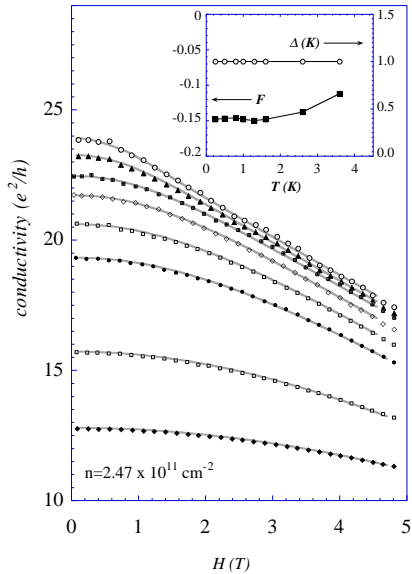}}
\caption{Conductivity of a Si-MOSFET versus in-plane magnetic field at
$T=0.25$, $0.5$, $0.8$, $1$, $1.3$, $1.6$, $2.6$, $3.6$ (K) (from top).
Symbols denote the experimental data; solid lines are calculated using
Eqs.~(\ref{all}), (\ref{ee}).  Inset: variation of
$F_0^\sigma$ and $\Delta_V$ with temperature obtained from fits to
Eq.~(\ref{ee}).  The electron density is $n=2.47 \times
10^{11}$ cm$^{-2}$.  The scattering time used to evaluate Eq.~(\ref{ee})
is $\tau = 5.28$ ps (see text for discussion).}
\label{fig1}
\end{figure}
}

The data in Fig.~\ref{fig1} demonstrate that there is very good
agreement between experiment and theory.  For
electron density $n=2.47\times 10^{11}$ cm$^{-2}$ a fit to the
theory Eq.~(\ref{ee}) of the magnetoconductivity data taken at
temperatures below $2$ K yields $F_0^\sigma=-0.15$ and $\Delta_V=1$ K.  At
temperatures above $2$ K we found an appreciable dependence of the
parameter $F_0^\sigma$ on temperature; all other electron densities
show similar behavior.

In Fig.~\ref{fig2} we show the in-plane magnetoconductivity for
different densities at a fixed temperature $T=1$ K. This temperature
corresponds to the linear portion of the conductivity vs. temperature
curve in zero magnetic field (see Fig.~\ref{fig3}), i.e. to the
ballistic regime $T\tau >0.1$ of the theory \cite{int,mpf}. The
comparison was done for magnetic field below $6$ T, where variations
of the magnetoconductivity are relatively small (well within the range
of applicability of the calculations). Again, the agreement between
theory and experiment is impressive. In the inset to Fig.~\ref{fig2}
we show the evolution of the fitting parameters $F_0^\sigma$ and
$\Delta_V$ with electron density. The Fermi liquid parameter
$F_0^\sigma$ is nearly independent of density, while the valley
splitting $\Delta_V$ fluctuates between $0.5$ and $1.2$ K.

{
\narrowtext
\begin{figure}[ht]
\epsfxsize=6.5 cm
\centerline{\epsfbox{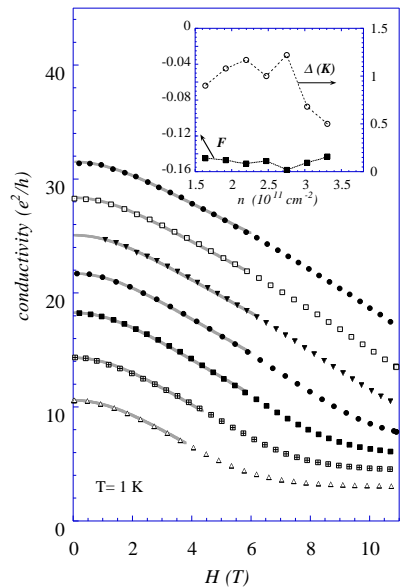}}
\caption{Conductivity of a Si-MOSFET versus
in-plane magnetic field for densities $n=3.3$, $3.0$, $2.75$, $2.47$,
$2.19$, $1.92$, $1.64\times 10^{11}$ cm$^{-2}$ (from top).
Symbols denote the experimental data; the
lines are calculated using Eq.~(\ref{ee}) (for fields
weaker than the field for full polarization).
Inset: $F_0^\sigma$ and $\Delta_V$ as a function of electron
density.  The temperature $T=1$ K.}
\label{fig2}
\end{figure}
}

We now compare the measured temperature dependence of the conductivity
in zero magnetic field with that predicted by the theory
\cite{int,mpf}.  In Fig.~\ref{fig3} we show the data together with the
predictions of the theory Eq.~(\ref{ee}) obtained using the values of
$F_0^\sigma$ and $\Delta_V$ determined from the magnetic field
dependence of the conductivity; see Fig.~\ref{fig2}. The variations of
the fitting parameters $F_0^\sigma$ and $\Delta_V$ with temperature
(see Fig.~\ref{fig1}), which were not taken into account in this
calculation, may change the high temperature tails of the calculated
curves at $T>2$ K in Fig.~\ref{fig3}. However, it is clear from
Fig.~\ref{fig3} that even for $T<2$ K, the temperature dependence of
the conductivity observed experimentally is considerably stronger than
that calculated from the theory \cite{int,mpf}. Thus, the parameters
deduced from fitting the field dependence of the conductivity do not
yield quantitative agreement with the observed temperature dependence.

{
\narrowtext
\begin{figure}[ht]
\epsfxsize=6 cm
\centerline{\epsfbox{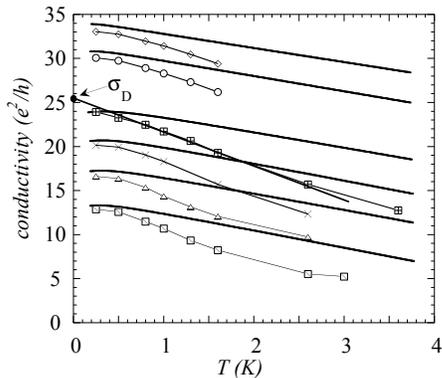}}
\caption{Temperature dependence of the conductivity of 2D electrons in
a Si-MOSFET at $H=0$ for densities $n=3.3$, $3.0$, $2.47$,
$2.19$, $1.92$, $1.64\times 10^{11}$ cm$^{-2}$ (from top).
The symbols indicate experimental data.  The solid lines are
calculated using the theory
\protect\cite{int,mpf}.  The straight line drawn for density $2.47
\times 10^{11}$ cm$^{-2 }$ is an extrapolation to zero temperature of the
(nearly) linear part of the conductivity used to determine the value of
$\tau = 5.28$ ps (see text).}
\label{fig3}
\end{figure}
}

We have attempted to fit the temperature dependence and the magnetic
field dependence simultaneously by allowing $F_0^\sigma$ and
$\Delta_V$ to vary freely with temperature.  Although a moderately
good fit can be obtained, the resulting variation of $\Delta_V$ with
temperature, although similar for different densities, is unacceptably
large (with variations up to a factor of $20$). Another possible
approach is to keep $\Delta_V$ small (as in Fig.~\ref{fig2}), to allow
$F_0^\sigma$ to assume the values needed to produce a good fit for the
temperature dependence of the conductivity, and to compare these
values with those in Fig.~\ref{fig2}. Clearly, the steeper
experimental curves in Fig.~\ref{fig3} correspond to larger absolute
values of $F_0^\sigma$, similar to those obtained in
Ref.~\onlinecite{kravch}.  Such disagreement between the values of
$F_0^\sigma$ obtained from two separate measurements illustrates once
more that the theory in its current form does not provide a consistent
description of all measurements in Si-MOSFET's. A possible explanation
of the above discrepancy is the neglect of inter-valley scattering in
the theory in its present form, or perhaps the presence of some other
source of scattering that has not been identified.

In summary, for the strongly interacting two dimensional system of
electrons in silicon MOSFET's, excellent agreement between experiment
and the theory of Zala, Narozhny and Aleiner \cite{int,mpf} was
obtained for the response of the conductivity to a magnetic field
applied parallel to the plane of the electrons. The Fermi liquid
parameter $F_0^\sigma(n)$ and the valley splitting $\Delta_V(n)$
obtained from fits to the magnetoconductivity, although providing
qualitatively correct behavior (including sign), do not yield
quantitative agreement with the temperature dependence of the
conductivity in zero magnetic field.  Our results suggest the
existence of additional scattering processes not considered by the
theory in its present form.

We are grateful to I.L. Aleiner, D. L. Maslov and B. Spivak for helpful
discussions. This work was supported by the U. S. Department of Energy at
City College under grant number DE-FG02-84ER45153 and at Brookhaven under
contract number DE-AC02-98 CH 10886.

\end{multicols}


\begin{references}
\bibitem{accuracy} The fits are more sensitive to $F_0^\sigma(n)$, which is
determined with an accuracy of $\pm 15$\%, than to the values chosen for
$\Delta_V$, which can be varied by $50$\%.
\bibitem{note}
Note that the description of transport and thermodynamic properties of
the 2DEG in terms of the single Fermi liquid parameter is an
approximation.  Strictly speaking, the constant that describes
the susceptibility is the zero$^{th}$ harmonic of the spin part of the Landau
function $f(\theta)$, while the constant that describes the conductivity
in the ballistic regime corresponds to backscattering,
i.e. $\theta=\pi$. The ZNA theory \cite{int} assumes the Fermi liquid
parameter to be independent of $\theta$ and thus uses just one
constant $F_0^\sigma$ to provide a phenomenological description of interaction
in the triplet channel.
\end{references}
\end{document}